# The Effect of Weekend Curfews on Epidemics: A Monte Carlo Simulation

Hakan KAYGUSUZ *

Department of Basic Sciences, Faculty of Engineering and Natural Sciences, Altınbaş University, Istanbul, Turkey.

Sabancı University SUNUM Nanotechnology Research Center, Istanbul, Turkey

hakan.kaygusuz@altinbas.edu.tr ORCID: 0000-0001-9336-1902

A. Nihat BERKER

Faculty of Engineering and Natural Sciences, Kadir Has University, Cibali, Istanbul, Turkey.

Department of Physics, Massachusetts Institute of Technology, Cambridge, Massachusetts 02139, USA

nihatberker@khas.edu.tr ORCID: 0000-0002-7868-5674

* Corresponding author (H. Kaygusuz)

Altınbaş University, Mahmutbey Dilmenler Cad. No: 26 Bağcılar Istanbul 34217 Turkey

hakan.kaygusuz@altinbas.edu.tr

+90 212 604 0100

**Abstract:** The ongoing COVID-19 pandemic is being responded with various methods, applying vaccines, experimental treatment options, total lockdowns or partial curfews. Weekend curfews are among the methods for reducing the number of infected persons and this method is practically applied in some countries such as Turkey. In this study, the effect of weekend curfews on reducing the spread of a contagious disease, such as COVID-19, is modeled using a Monte Carlo algorithm with a hybrid lattice model. In the simulation setup, a fictional country with three towns and 26,610 citizens were used as a model. Results indicate that applying a weekend curfew reduces the ratio of ill cases from 0.23 to 0.15. The results also show that applying personal precautions such as social distancing is important for reducing the number of cases and deaths. If the probability of disease spread can be reduced to 0.1, in that case the death ratio can be minimized down to 0.

**Key words**: Monte Carlo simulation, epidemic, curfew, SIQR, COVID-19

**1. Introduction**

The Coronavirus disease 2019 (COVID-19) pandemic caused by the severe acute respiratory syndrome coronavirus 2 (SARS-CoV-2) is currently ongoing since December 2019. To the best of our knowledge, there is no approved specific antiviral drug treatment for SARS-CoV-2 (Tarighi et al. 2021) and vaccination efforts are conducted worldwide. The spread of the disease is ongoing and various countries are applying nationwide and/or local measures for preventing the spread of the disease.

As of May 2021, different countries and communities follow different ways to combat the disease: Vaccination, partial curfews and full quarantines. Among the examples, Turkey applies a partial curfew which applies a travel ban for all citizens (except for mandatory duties) on weekends. This 2-day curfew is applied at least to the end of May 2021.

By the means of in silico and computational methods, many scientists are working on discovering the genome analysis and mutations (Ugurel et al. 2020, Eskier et al. 2021), using epidemiological simulation. By the means of epidemiological data fitting and simulation there are two main approaches: The first one is to fit the data with mathematical models and the second one is to simulate and generate data in consistency with the real clinical data (Maltezos and Georgakopoulou 2021). Previously, several studies on modeling infectious diseases by the means of random walk and stochastic processes have been reported (Filipe and Gibson 1998, 2001, Draief and Ganesh 2011, Bestehorn et al. 2021). Monte Carlo (MC) simulation is one of the efficient methods for generation of a decision making tool (Xie 2020). Different MC studies have been reported for COVID-19 spread, such as analyzing different scenarios for selected countries (Vyklyuk et al. 2021), age-structured mobility data for simulation of the pandemic spread in selected cities (De Sousa et al. 2020) and random-walk proximity-based infection spread (Triambak and Mahapatra 2021).

The aim of this paper is to observe the effects of disease detection rate and the chance of recovery for a contagious disease under a weekend curfew scenario, similar to the one applied in Turkey. In the model, three different towns with a total population of 26,610 were modeled in a hybrid model of susceptible, infected, quarantine and recovered (SIQR) model (Kermack and Mckendrick 1927, Huppert and Katriel 2013, Shu et al. 2016, Cooper et al. 2020, Odagaki 2020, Odagaka 2021), lattice model (Liccardo and Fierro 2013) and spin-1 Ising model (Berker and Wortis 1976, Hoston and Berker 1991).

**2. Methods**

The Monte Carlo model used in this study is written in the C & C++ language by the authors. A fictional country with a population of 26,610 and three separate towns (namely A, B and C, as shown in Figure 1) are generated. The following assumptions were made in the simulation: Town A is the biggest, followed by Town C and Town B, all towns have the same population density of 0.89 and the towns are reachable by airports (each town has one airport). Each person in the towns is placed randomly on an Ising-like model where the neighbors of each particle can be another person or a void space (as shown in Figure 2). In order to improve the randomness in movement, there are also "obstacles" which are simply lattice points which are impossible to be on.

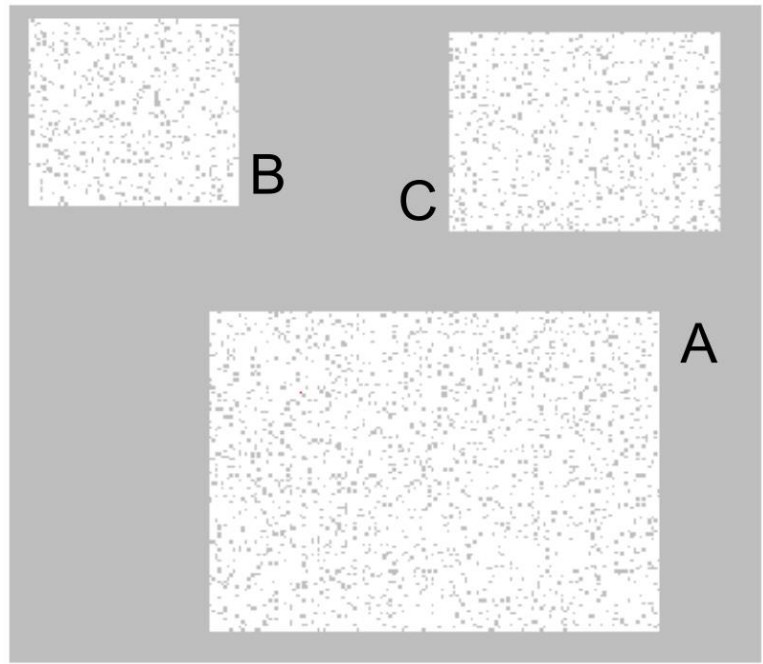

**Figure 1** Three fictional towns A, B, C which hold 26,610 persons in this study.

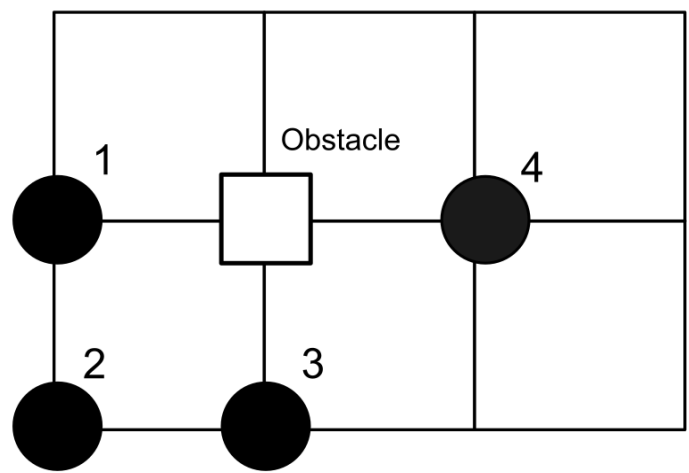

**Figure 2** An example to the four of the particles in the given lattice, where each particle represents a person in the simulation. The obstacle is a forbidden lattice point.

The following attributes are possible for each particle: a) Healthy, b) sick and undiagnosed, c) sick and in quarantine, d) recovered (cured), e) dead. Option b "sick and undiagnosed" is used for simulation of both undiagnosed persons and asymptomatic cases.

In each Monte Carlo step, particles are selected, and a random possible direction is assigned to this particle as the movement. After selecting and checking the movement of all possible particles, this is called one Monte Carlo step and named as "day" in the simulation. In the following "day", the process is repeated for all particles.

On day 7, a randomized person gets infected on Town A and starts spreading the disease (patient zero). On day 15, the disease spreads to a random position in Town B. On day 22, the disease spreads to a random position in Town C.

An infected person can spread the disease at a certain probability $P_{spread}$ and the sickness is detected at a probability of $P_{detection}$.

If the person is diagnosed with the disease, then the rate of successful treatment is $P_{treatment}$. This person goes to quarantine, which means the person is immobile and cannot transmit the virus ($P_{spread} = 0$) until the disease is cured.

If the disease in the person is not properly diagnosed, then the rate of the survival is $P_{survival}$. In both cases, the person either gets cured or dies (based on $P_{treatment}$ or $P_{survival}$) after 28 days.

The government applies a curfew on weekends like Turkey, beginning on day 13, which is 6 days after the first detection. On curfew, persons are immobile and are not able to spread the disease. On weekdays (5 days), all persons are mobile and can spread the virus except the persons under quarantine. A cured person cannot be reinfected and does not spread the virus. This model assumes that all persons respect the weekend curfews and no violations occur. Additionally, the workers in mandatory services such as health and security are neglected.

Spread of the virus and infection is modeled by the generation of a random number r on each Monte Carlo step. If $r > P$, then the aforementioned event occurs. For instance, if day $> 28$ and $r > P_{treatment}$, then the person gets recovered.

## 3. Results and Discussion

Figure 3 shows the spread of the disease on days 0, 11 and 24. The ill persons are marked with red, indicating the random movement of the ill persons as well as the spread of disease at a probability of $P_{spread}$.

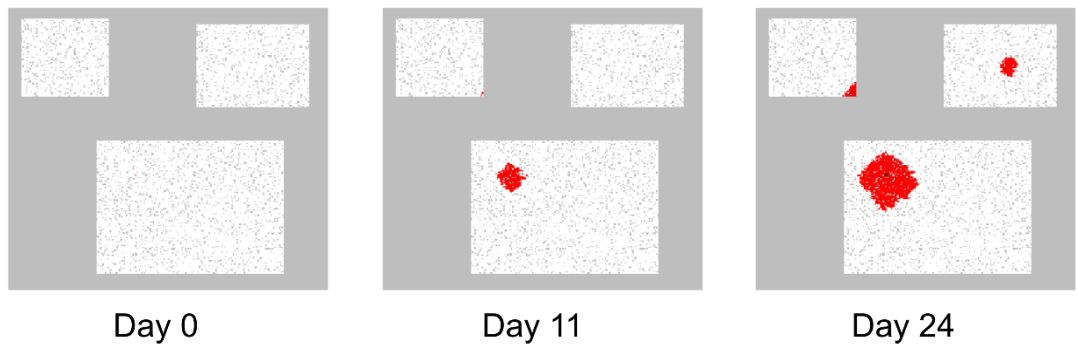

Day 0        Day 11        Day 24

**Figure 3** Increase in the number of ill persons on three different days, which includes both the spread of the disease and movement of the persons.

To test the model, a SIQR - susceptible, infected, in quarantine and recovered – plot is generated as shown in Figure 4. SIQR model is used for modeling COVID-19 as shown in literature (Odagaki 2020, Odagaki 2021). In the test model, nobody dies because of the disease and accordingly, with the onset of the infection, the sums of the ratios of each case are always 1.0, indicating the model works for the given simulation setup.

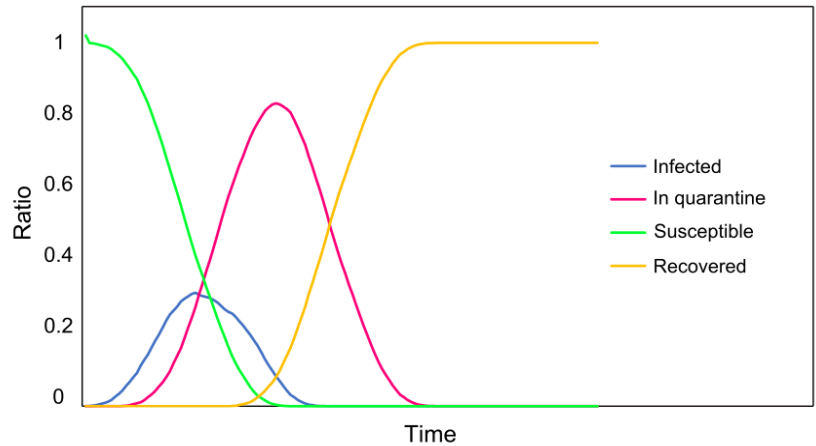

**Figure 4** Ratios of susceptible, infected, in quarantine and recovered in this model.

The effect of weekend curfews is shown in Figure 5. This figure indicates the start day of the weekend curfews when compared to the case without the weekend curfews, where all plots were obtained at a detection probability of 0.3 for comparison. As indicated in the plot, starting the weekend curfew early as day 14 decreases the ratio of ill persons from 0.22 to 0.15, as well as widening the peak. Reducing the amount of ill persons at a given time is very important for reducing the workloads of health professionals and efficiently using capacities of intensive care units (Farsalinos et al. 2021). Therefore, the effect of curfews are clearly visible in the given simulation. Prolonging the starting day of the weekend curfews slowly shifts the curve to the left, i.e. to the case without any curfew/nationwide lockdown. Another interesting result is the slight extension in the duration of the epidemic when weekend curfews are applied. This is due to the nature of mobility restriction, for the case of non-restrictions the virus will spread easily, and the immunity will be reached earlier. Of course, one should note that reaching herd immunity is not the main goal for many health systems, primary goal is the reducing the death rate and workload of health facilities.

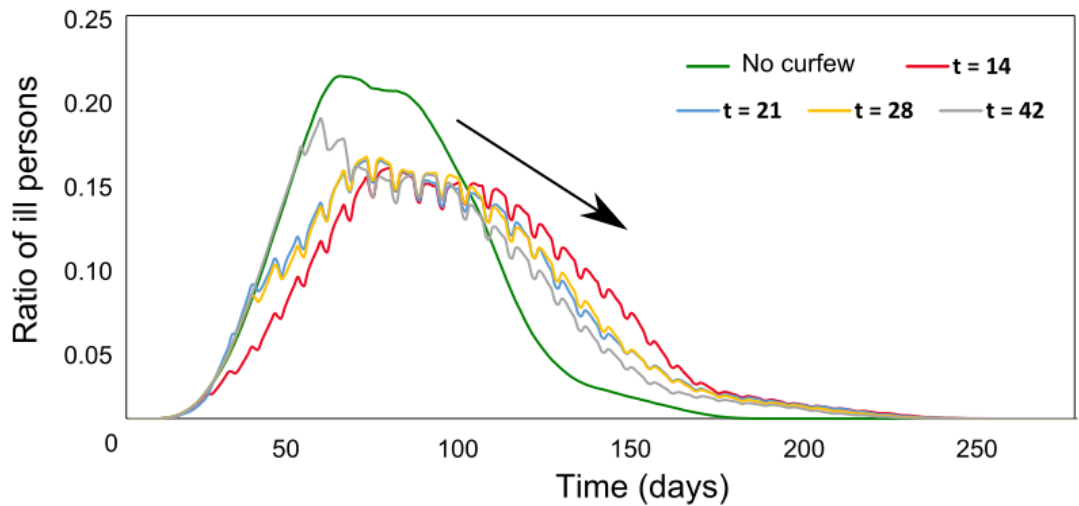

**Figure 5** Ratio of ill persons for five different cases: No curfew is applied (green), weekend curfews starting on day 14 (red), day 21 (blue), day 28 (yellow) and day 42 (gray).

In the second part of the study, the effect of varying $P_{spread}$ on the death rate was investigated. Here, $P_{spread}$ varied from 0.1 to 1.0, with an increase of 0.1 and the simulation repeated for different $P_{spread}$ values, with an assumption of asymptomatic cases always recover, where weekend curfews continue. The result of this simulation is shown in Figure 6. Reducing the $P_{spread}$ decreases the death ratio significantly, which demonstrates the importance of social distancing, wearing masks and complying with other hygiene precautions in combating the spread of the disease.

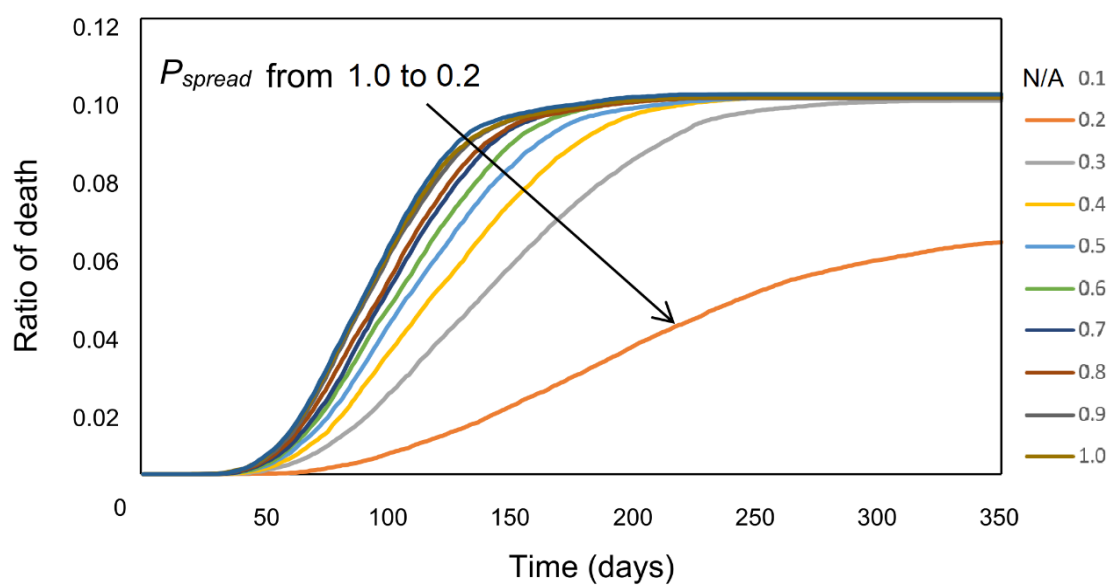

**Figure 6** Effect of the spread probability P$_{spread}$ on the death ratio, where decreasing the probability gives the lower ratio (orange curve). P$_{spread}$ = 0.1 is not shown since the ratio is ~0.

Figure 5 and Figure 6 are among the key findings of this study. Most important result is that the early start date of curfews are more effective than later start dates by the means of ratio of ill persons, and it is very effective when compared to the case without curfew application (ratio of ill persons is reduced from ~0.23 to ~0.15). Other result is the effect of the *P$_{spread}$*. This can be read in many ways, such as applying the social distancing and it is clearly visible that the death rate is decreased by half by reducing the ratio to 0.2 (Figure 6). By reducing to 0.1, the death ratio almost becomes 0, which is an important indicator for measures such as effective social distancing, hygiene precautions and even vaccination.

In the world, especially in the east Africa, there were some obstacles to COVID-19 control. For instance, some countries used the curfews and/or partial lockdowns, in some others there were problems that prevented an effective solution (Nakkazi 2020). There are several reports on the impact of curfews on the recent COVID-19 pandemic. Among the reports, one of the important study is the effect of curfews on SARS-CoV-2 transmission in French Guiana (Andronico et al. 2021). French Guiana applied a curfew in 2020. The major result is the reduction in the basic reproduction number of the virus from 1.7 to 1.1 and this reduction was enough for avoiding the hospital saturation (Andronico et al. 2021).

For the cases of Germany and Switzerland, a study indicates that an earlier lockdown is more effective than a later one for COVID-19 prevention (Huber and Langen 2020) and that is similar to our result in Figure 5, where an earlier start date is more effective in decreasing the number of infected persons. Turkey is also among the countries which applied the curfews in COVID-19 outbreak control, which started on April 11, 2020 (Demirbilek et al. 2020). This restriction is later lifted especially after the vaccination process become widespread for the population.

In further studies, the following parameters can be investigated in the simulation: Including age and risk groups in the population, the effect of urbanization, income, remote working ability, and education levels. For each case, proper mobilization, spread, and recovery rates should be assigned since different groups can have different levels of access to health and mobility services. and the simulation should be set accordingly.

**Conclusion**

By applying a weekend curfew e.g., two-day alternating lockdowns and five-day free motion on particles being simulated in this study, it is found that the ratio of infected decreases by 68% at the peak point. It is also evident that starting this curfew is more efficient when started earlier (day 14 in this study). Turkey also observed a similar pattern of case decrease by following the weekend curfews for combating COVID-19. Monte Carlo simulations remain as a strong tool not only for predicting the spread of contagious diseases, but also for modeling the alternative precaution measures. Further and detailed studies might show the effects of disobeying the curfew, age, gender and other parameters as well as the effect of vaccines altogether for modeling the epidemic/pandemic.

**Acknowledgments**

A. Nihat Berker gratefully acknowledges support by the Academy of Sciences of Turkey (TÜBA).